\documentclass[doublecol]{epl2} %for 2 columns style without line numbers
\usepackage{graphicx}

\title{Step energies and equilibrium shape of strained monolayer islands}

\author{J.E. Prieto\inst{1} \and I. Markov\inst{2}}

\institute{
  \inst{1} Centro de Microan\'alisis de Materiales, Departamento de F\'\i{}sica de la
Materia Condensada, Condensed Matter Physics Center (IFIMAC) and Instituto
``Nicol\'as Cabrera'', Universidad Aut\'onoma de Madrid, E-28049 Madrid, Spain\\
  \inst{2} Institute of Physical Chemistry, Bulgarian Academy of Sciences,
1113 Sofia, Bulgaria 
}

\abstract{
Using a simple atomistic model of anharmonic nearest-neighbors interaction,
we have calculated the step energies of strained hexagonal monolayer
islands. These have been found to decrease with the absolute value of the misfit
due to the strain relaxation at steps. The effect is significantly more
pronounced in the case of positive misfit owing to the stronger repulsive
interatomic forces. Furthermore, (111)-faceted steps are favored at positive
misfit (compressed islands) and, to a lesser extent, (100)-faceted steps at
negative misfits (tensile islands). The result is rationalized in terms of the
different bonding geometries at step edges and a comparison with experiments is 
included. Thus, the equilibrium shape
transforms from regular hexagons at zero misfit to threefold symmetric hexagons
with increasing misfit.
}

\begin{document}

\maketitle

The problem of the equilibrium shape of crystals has a long and venerable
history~\cite{Gib78,Cur85,Wul01,Her52,Lan58} 
(for a review see Ref.~\cite{Bon03}). 
In recent decades this problem in connection with
epitaxial stress has attracted much attention in the case of the formation
of self-assembled quantum dots~\cite{Ter93,Ter94,Mul00,Pol00}. The equilibrium
shape of two-dimensional monolayer-high islands has also emerged as an
important topic in connection with facet-shape changes of three-dimensional 
crystallites~\cite{Bon03}, the possibility of measuring the absolute values of
the step and kink formation energies~\cite{Gie01} and the growth of crystal
surfaces~\cite{Bur51}.

The initial stages of the epitaxial growth of thin films include the formation
and growth of elastically-strained monolayer islands~\cite{Mar87,Pri95}. 
This has motivated an interest in the study of their equilibrium
shapes~\cite{Li00,Pra04,Feng06,Zan07}. These authors have addressed the behavior 
of the total energy of rectangular islands, considered to be given by the
sum of strain and step energies, the latter being assumed as
strain-independent~\cite{Li00,Pra04,Feng06}. The systems studied are characterized
by a highly anisotropic surface stress, as for example, reconstructed, single-domain
Si(100) terraces.  
But also for the case of isotropic epitaxial strain, authors find an 
island shape instability via long-range step-step interaction even in the absence 
of step-energy anisotropy~\cite{Li00,Pra04,Feng06}.

\begin{figure}[h,t]
\center
\includegraphics*[width=7.5cm]{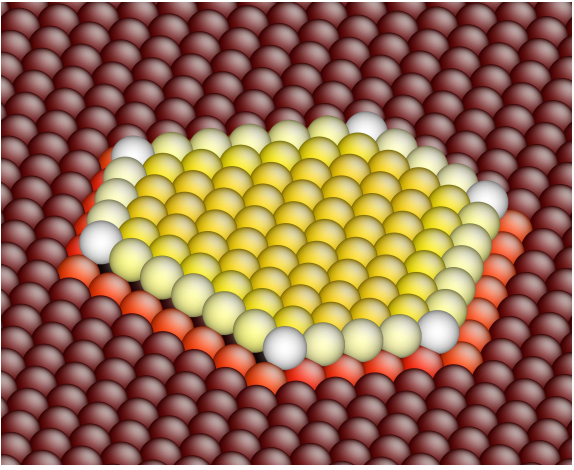}
\caption{\label{balls} Hexagonal, threefold symmetric island of 5 atoms in
the (111)-faceted, $A$ steps and 7 atoms in the (100)-faceted, $B$ steps (5,7).
The color scale denotes the height of the considered atom and has
been represented using the AtomEye software~\cite{Li}. This height is
measured above the level of the corresponding crystallographic plane, but
a constant fraction of the interlayer distance has been added in order to
better distinguish atoms from different levels. The height is biggest at
edges and corners due to the atoms ``climbing up" on their neighbours
underneath due to strain relaxation. The lattice misfit is -7\%.
}
\end{figure}

On the contrary, in the present Letter we consider islands on highly
symmetric sixfold coordinated surfaces, 
for which the surface stress tensor is isotropic~\cite{Mch80} 
and we set out to study the contribution of 
strain, not considered before, 
to the step energies and in turn to the aspect ratio. 
We chose to use the simplest model that contains the essential 
ingredients to describe the effect of strain.
We perform atomistic calculations using model 
potentials for simulating hexagonal islands at $T=0$~K. 
Our approach follows the spirit of the work of Hamilton on overlayer 
strain relief in {\em fcc}(100) surfaces~\cite{Ham02}.
As known, hexagonal islands are enclosed by steps of two different 
types: (100)-faceted or $A$ steps and (111)-faceted or $B$ steps which, 
when circulating along the island perimeter, follow the alternating 
sequence $BABABA$ (see Fig.~\ref{balls}).
In these highly symmetric surfaces, equilibrium island shapes are 
expected to be  largely determined by the ratio 
$r = \gamma_{111}/\gamma _{100}$ of the free energies of $B$ and $A$ 
steps. For some homoepitaxial systems, this ratio has been studied
in detail (results are summarized in Ref.[\cite{Gie01}]). 
In general $r$ differs from unity. A relatively large deviation 
($r^{-1} = 1.15$) has been reported for vacancy islands on Pt(111) 
by Michely and Comsa~\cite{Mich93} from the measured ratio of the 
step lengths.

We consider an atomistic model of a rigid substrate with sixfold symmetry and
lattice parameter $a$ that can represent {\em fcc}(111) or {\em hcp}(0001)
surfaces on which monolayer islands grow. Atoms interact in the first
coordination sphere through an anharmonic Morse potential~\cite{Mar93}
\begin{eqnarray}\label{potent}
V(r) = V_{0}[e^{-12 (r-b)} - 2e^{-6 (r-b)}],
\end{eqnarray}
where $b$ is the equilibrium atom separation, so the lattice misfit is
given by $\varepsilon = (b-a)/a$. For every configuration, the cluster is
allowed to relax by varying iteratively the atomic positions until all the
forces fall below some negligible cutoff value. Clearly, our atomistic model is
intentionally oversimplified in order to study the effect of the lattice misfit
in its pure form. It obviously gives a regular hexagon, $l_{100}=l_{111}$, as the
equilibrium island shape at zero misfit.

\begin{figure}[h,t]
\center
\includegraphics*[width=8.5cm]{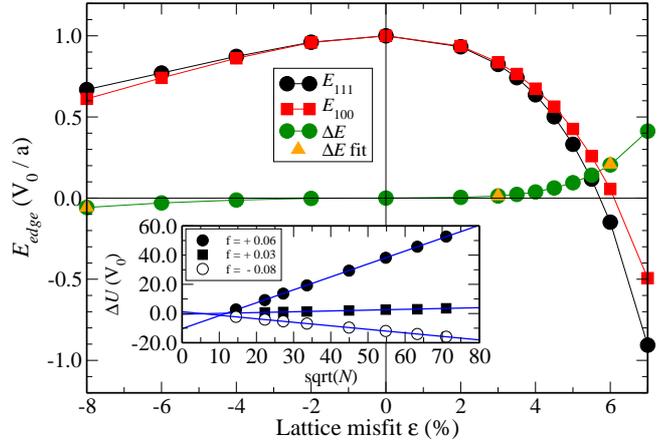}
\caption{\label{E_edge} Dependence of the energies of $A$ and $B$ steps,
in units of $V_0/a$, on the lattice misfit. The third curve represents the
difference of the step energies. The triangular points are 
the values calculated from triangular islands, as shown in the insert.
Insert: Difference of the total energies of triangular
islands, in units of $V_0$, as a function of the square root of the total
number of atoms. These islands are enclosed by only either $A$ or
$B$ steps depending on the orientation. Results for three values of the 
lattice misfit are given: 3, 6 and -8\%.}
\end{figure}

We calculate the values of the energies of $A$ and $B$ steps by means of the
following procedure. 
The total energy of a strained polygonal island contains contributions from 
the interior bonds, which are the most highly strained due to the lattice 
misfit, and from the edges and corners, where strain is relaxed. 
In order to compute the step energy of a regular-hexagonal island of
($n, n$) atoms at the edges (where the first and second numbers denote
the number of atoms in the $B$ and $A$ edges, respectively) we consider the
two configurations ($n+1,n$) and ($n-1,n$). We then subtract the
energies of the two islands and correct for the different number of atoms contained
by adding the energy of the corresponding number of completely strained ``interior"
atoms at the center, a contribution that can be calculated analytically from the
known interaction potential, [Eq.~(\ref{potent})]. The corner energies can be
assumed to be the same in both islands if they are large enough, so that the
energy difference gives the increase in edge energy due to the increase of the
length of the $B$ steps. The edge energy of $A$ steps can be calculated
analogously by comparing ($n,n+1$) and ($n,n-1$) islands. It is clear that in
this way we are considering all effects related to strain relaxation at edges
to be included into the step energy. This makes sense since otherwise the only 
contribution to the step energy would be the number of broken bonds which,
being the same for $A$ and $B$ steps, would result in 
no difference in their edge energies. The results of these calculations 
are shown in Fig.~\ref{E_edge}.

Since this is a point of crucial importance, we now proceed to check our
assumption that strain affects the edge energies. In order to eliminate the
effect of the ``interior" strain, we compare the total energies of {\em triangular}
islands of the same size, but enclosed by either only $A$ or only $B$ edges. 
Because of their equal sizes, both types must contain equal contributions 
of ``interior" strain and can only differ in the way strain is relaxed at edges 
of different types. The energy difference of such islands is then given by
\begin{equation}
\Delta U = 3\Delta E_{corner} + 3[(\sqrt{2N}-1/2)-2l_c]\Delta E_{edge}
\end{equation}
where $\sqrt{2N}-1/2$ is the full length of the edge expressed as a function
of the total number $N$ of atoms in the island (here a factor 1/4 has been
neglected compared to 2$N$), the term in square brackets gives an
``effective" length of the edge by subtracting an extension $l_c$ around each
corner where edge energies are affected and finally, 
$\Delta E_{edge}=\gamma_{100} - \gamma_{111}$ is the difference in energies
between the two types of steps.

The insert of Fig.~\ref{E_edge} shows the difference of the energies of triangular
islands enclosed by either $A$ or $B$ steps at three different values of the
misfit, 3\% and 6\% (compressed islands), and -8\% (tensile islands). The data can
be very well described by straight lines, confirming the validity of the
considerations above. From the slopes, the difference in edge energies
$\Delta E_{edge}$ can be calculated. The slope is positive for compressed islands
($\gamma_{100} > \gamma_{111}$) and negative for islands under tensile stress
($\gamma_{100} < \gamma_{111}$), in agreement with Fig.~\ref{E_edge}.
Furthermore, the result is in very good quantitative agreement with the value
obtained by subtracting the values of $\gamma_{100}$ and $\gamma_{111}$
calculated directly, as shown in Fig.~\ref{E_edge}, giving consistency to our
procedure. In addition, to the extent that the difference $\Delta E_{corner}$ in
corner energies between the two types of triangular islands can be neglected,
the extension $l_c$ of the region at the step affected by the corner can
be calculated from the $y$-intercept of the lines in the insert of Fig.~\ref{E_edge}.
We obtain the values of 5.6 and 8.0 atomic spacings for 3\% and 6\% misfit,
respectively, and 3.4 spacings for -8\%. These results appear very reasonable,
of the order of a few lattice spacings, 
increasing with misfit and larger for positive than for negative values of the 
misfit, because of the more efficient relaxation at edges and corners~\cite{Mar84}.

Figure~\ref{E_edge} demonstrates our main result, the behavior of the step
energies, $\gamma_{100}$ and $\gamma_{111}$, as a function of misfit. A number
of relevant facts can be observed. The energy of both types of steps has a
value of $V_0/a$ at zero misfit in accordance with dangling-bond counting.
We see that both edge energies decrease with increasing absolute
value of the lattice misfit, as expected due to strain relaxation at island
edges, but the effect is much stronger in the case of compressive misfits
because of the more effective relaxation: The region with relaxed bonds extends
significantly longer into the island in the case of positive
misfits~\cite{Korfu}. Another point, which will be addressed below, is that
the energies become negative for high enough positive values of the misfit.

A very important feature visible in Fig.~\ref{E_edge} is the {\em opposite
sign of the difference between the step energies} of $A$ and $B$ steps. For
compressed islands, i.e. for positive values of the misfit, $B$-type,
(111)-faceted steps are significantly favored, while $A$-type, (100)-faceted
ones have lower energy for negative values of the misfit, although the effect 
is smaller. So (111)-faceted steps will be favored for positive and
(100)-faceted steps for negative misfits. This is a clear prediction of our
model and can be rationalized as follows: A compressed island relaxes strain at
steps by the displacement of edge atoms (from the bottoms of the corresponding
potential troughs they would occupy in the absence of misfit) 
in the outward direction~\cite{Fra49}.
For (111)-faceted steps, this means that edge atoms are pushed 
towards ``bridge" positions of the substrate layer
underneath, while for (100)-faceted steps, they are pushed towards ``on-top"
positions (see Fig.\ref{balls}). Sensibly, the former situation has a lesser cost
in bonding
(adhesion) energy than the latter. The situation is clearly reversed in the case
of negative misfits, i.e. in tensile islands, for which strain relaxation at
steps forces edge atoms to be displaced in the inward direction. Now in this case, 
atoms at (111)-faceted steps are pushed towards ``on-top" positions and atoms 
at (100)-faceted steps towards less unfavorable ``bridge" positions.

\begin{figure}[h,t]
\center
\includegraphics*[width=8.5cm]{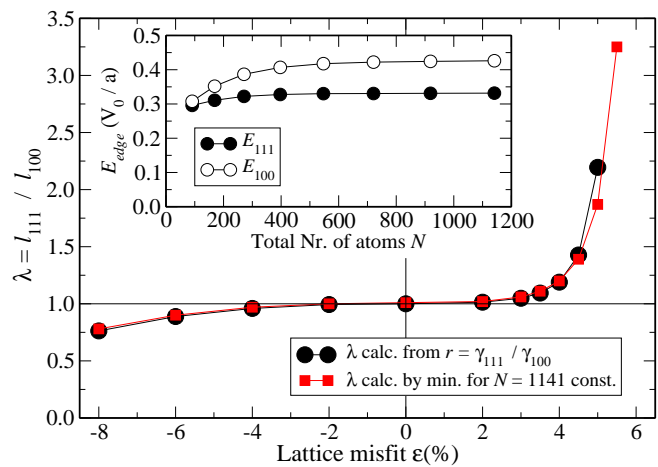}
\caption{\label{Shape} Dependence of the ratio of the step lengths (the
equilibrium shape) of $A$ and $B$ steps on the lattice misfit, calculated
both from the differences in edge energies obtained from Fig.~\ref{E_edge} and
from a direct search of the minimum energy among islands of nearly constant number of
atoms (see text). Insert: Dependence of the energies of $A$ and $B$ steps on
island size.}
\end{figure}

Figure~\ref{Shape} gives the aspect ratio of the step lengths
$\lambda$ = $l_{111}/l_{100}$, characterizing the equilibrium shape of the
islands, as a function of the misfit. Data in this plot have been determined in
two different ways: First, from the calculated values of the step energies
$\gamma_{111}$ and $\gamma_{100}$ (shown in Fig.~\ref{E_edge}), the aspect 
ratio $\lambda$ has been obtained by minimization of the perimeter energy at 
constant number of atoms (island area), as given by
\begin{equation}
\label{rvsgamma}
\lambda = \frac{r - 2}{1 - 2 r},
\end{equation}
where $r = \gamma_{111}/\gamma_{100}$ is the ratio of the step energies,
obtained from Fig.~\ref{E_edge}. The second procedure is the direct calculation
of the total energy of islands of different lengths $l_{100}$ and
$l_{111}$ but with constant total number of atoms to find the configuration of
minimum energy. Since this is not possible to achieve exactly for a large enough
number of different configurations with integer number of atoms, 
several ones containing
numbers very close to the desired size (e.g. 1141 atoms) were selected and the
minimum-energy configuration was determined by interpolation.
Figure~\ref{Shape} shows that the two procedures give basically the same results,
confirming the validity of the approach. We see that the $(111)$-faceted step becomes
several times longer that the $(100)$-faceted one at positive misfit. The opposite
tendency is observed for tensile islands, although the effect is smaller.

The question of the negative edge energies at high positive values of the misfit
can now be addressed. Considering Eq.~(\ref{rvsgamma}), it is immediately seen
that for $r \rightarrow$~1/2, $\lambda$ tends to infinity, meaning that the
threefold symmetric hexagon has transformed into a triangle exposing only
$B$-steps. An inspection of Fig.~\ref{E_edge} shows that negative edge energies
appear for misfits higher than about 5.5\% in our simplified model. For those
values of the misfit, Fig.~\ref{Shape} predicts already a factor $\lambda$ of
the order of~3. This means that negative edge energies, which would imply an
instability towards disintegration of large islands into many small islands, are
reached at very high values of the misfit, together with a clear preference
for a given type ($B$) of steps. For those values, real systems will probably relax
strain by some mechanism not considered in our model, such as the introduction 
of misfit dislocations.

\begin{figure}[h,t]
\center
\includegraphics*[width=7.5cm]{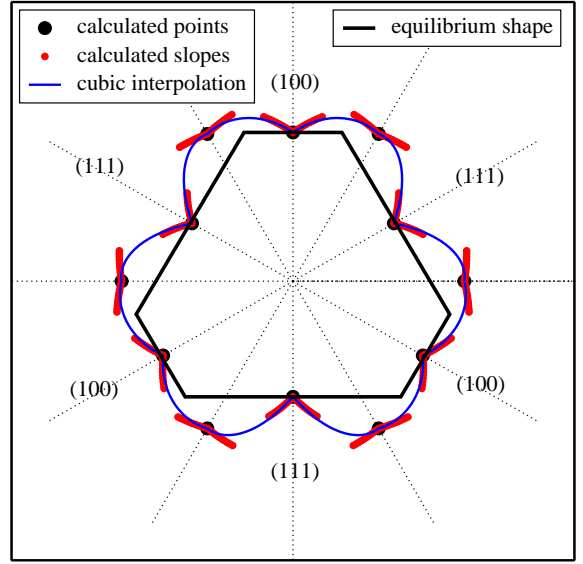}
\caption{\label{wulff} Polar plot $\gamma (\theta)$ and Wulff's
construction of an island under compression of 5\%.}
\end{figure}

The insert of Fig.~\ref{Shape} shows the dependence of the edge energies on 
island's size.
As seen the energy is size-independent with the exception of very small sizes
where the elastic fields at the edges substantially overlap and the contribution
of corners becomes significant. This is in accordance with the interpretation of 
the edge energy as a macroscopic concept. Finally, Fig.\ \ref{wulff} shows the
polar plot $\gamma (\theta)$ together with
the Wulff construction for the determination of the equilibrium shape of an
island compressed at 5\%. For this plot, in addition to the edge energies
$\gamma_{111}$ and $\gamma_{100}$ (and the corresponding 
$\gamma_{\overline{1}13}$, which does not appear in the equilibrium shape) 
the kink energies $\beta$ in those steps were calculated in order to obtain through
\begin{eqnarray}
\label{steptheta}
%\nonumber
\gamma(|\theta|) = \gamma(0)\cos(|\theta|) + \frac{\beta}{a} \sin(|\theta|),
\end{eqnarray}
in complete analogy with the three-dimensional polar plot~\cite{Che84},
the dependence of the step energy on orientations close to that of a smooth, 
kinkless step (of energy $\gamma (0)$) with its corresponding singular 
point in the $\gamma (\theta)$ polar plot. 

\begin{figure}[ht]
\center
\includegraphics*[width=8.5cm]{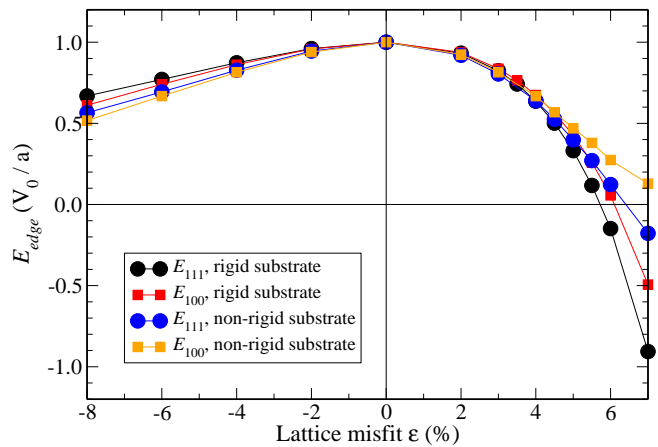}
\caption{\label{S1} 
Dependence of the energies of $A$ and $B$ steps,
in units of $V_0/a$, on the lattice misfit, calculated by allowing atoms in the topmost
atomic layer of the substrate to relax in the same way as the atoms in the island, 
as explained in the text. In addition, the data for the rigid substrate shown in 
Fig.~\ref{E_edge} are included for comparison.} 
\end{figure}

The most important conclusion of our work is that compressed islands are 
predicted to possess an equilibrium shape containing longer (111)-faceted steps. 
On the contrary, tensile islands
will show a nearly regular hexagon as the equilibrium shape, with a slight
preference for (100)-faceted steps, as the step energy difference is very small.
Concerning the limitations of our approach, it is important to be aware that 
we are considering only the effect of strain in the equilibrium shape of
epitaxial monolayer islands on hexagonal substrates. Therefore, our model is
best suited for the description of simple metallic systems. 
However, other aspects neglected in our simulations are found to play only minor roles
determining equilibrium shapes in our conditions. 
For example, the assumption of a rigid substrate was
checked in simulations allowing a number of layers of the substrate to relax.
Figure~\ref{S1} shows the energies of $A$- and $B$-type steps calculated by allowing 
one monolayer of the substrate to relax in the same way as the atoms in the island, 
together with the results for a rigid substrate shown in Fig.~\ref{E_edge}. 
The result is that the difference in energies of the two type of edges is very similar in
both cases. We conclude that the contribution of elastic interactions mediated by the 
substrate does not significantly influence equilibrium shapes for the relatively 
large island sizes and the highly (threefold) symmetric substrate considered here.
Also the assumption of different bonding strengths between island
atoms (cohesion) and between island and substrate atoms (adhesion) does not significantly
affect island shapes, as was checked by simulations for different values of the adhesion
parameter between island and substrate. This confirms that our simulations contain the
essential ingredient determining equilibrium shapes, the different lattice parameters
of film and substrate materials. 
Clearly, equilibrium shapes can be affected by several additional factors not
considered here, such as the presence of dislocation networks, surface 
reconstructions, stacking faults, interdiffusion with the substrate, changes 
in stoichiometry of compounds, etc. These factors also explain the scarcity of 
reliable examples of experimental results to compare with. Some of them are 
discussed in the following. 

\begin{table}[h] \centering
%\begin{table}[htbp] \centering
\begin{tabular}{|c|c|c|c|} \hline
System [Ref.] & Strain & Shape & Comm. \\
              &        & Pref. facet & \\ \hline \hline
Ag/Pt(111)~\cite{Rod93} & 4.3\% & 3-f hex. & \\
                        &       &    (111) & \\ \hline
Co/Pt(111)~\cite{Bru08} & $>$ 0,  & compact & recons. \\ 
                        & recons. & (111)  & \\ \hline
Co/Ru(0001)~\cite{FEG07} & -7.3\% & triang. & poss. \\
                         &        & (100)   & kinet.\\ \hline
Cu/Cu(111)~\cite{Gie01} & $\sim$ 0 & $\sim$ 6-f hex. & \\ \hline
Ag/Ag(111)~\cite{Gie01} & $\sim$ 0 & $\sim$ 6-f hex. & \\ \hline
Au/Au(111)~\cite{Mich93} & $>$ 0,  & 3-f hex. & recons.\\
                         & recons. &  (111)   & \\ \hline
Pt/Pt(111)~\cite{Mich93} & $>$ 0, close  & 3-f hex. & poss.\\
                         & to recons.    & (111)    & recons.\\ \hline
\end{tabular}
\caption { \label{tbshapes}
Compilation of experimental results on equilibrium shapes of epitaxial 
systems with threefold or sixfold symmetry for which a reasonable comparison with the
results of our simulations can be performed, as discussed in the text. Abbreviations
used: Pref.: Preferred; Comm: Comments; 3-f: threefold; 6-f: sixfold; hex.: hexagonal; 
triang.: triangular; recons: reconstructed; 
poss.: possibly; kinet: kinetics. 
The systems Co/Pt(111), Au/Au(111) show a positive surface strain due to surface
reconstructions in which additional rows of atoms are inserted. The Pt(111) surface
is also prone to a similar surface reconstruction at the growth temperatures 
required for achieving equilibrium shapes. See text and cited References for details.
}
\label{tb:values}
\end{table}

Clear equally oriented threefold hexagonal islands with longer (111)-faceted steps
have been reported as the equilibrium shape in the epitaxial system Ag/Pt(111)
with a compressive stress of 4.3\%~\cite{Rod93}. Authors calculate a ratio of step
free energies $\gamma_{100} / \gamma_{111}$ =~1.25, which gives a ratio of edge
lengths of about 2, in agreement with our model (see Fig.~\ref{Shape}).
Another interesting system is Co/Pt(111). Here, if deposition is performed at low
temperatures to avoid interdifussion, quasi-hexagonal, monolayer islands
are obtained as equilibrium shapes upon annealing~\cite{Cren04,Bru08}. These 
islands show a network of partial dislocations on top as a result of an increase of
the density of Co atoms, which turns the stress from tensile to compressive in the
Co monolayer and allows the island to expose only (111)-faceted steps in the
equilibrium shape~\cite{Bru08}, in agreement with our prediction.

In further different metallic epitaxial systems, triangular islands have been
found, but these most probably do not correspond to thermodynamic equilibrium shapes
and are determined instead by kinetics (e.g. different diffusion coefficients along
different step types), although both facts might be related. Examples are
Co/Ru(0001)~\cite{Hwa92} and Co/Cu(111)~\cite{Fig93}.
For the tensile Co islands on Ru(0001), a careful structural determination by
combined low-energy electron diffraction and microscopy has shown that the
preferred microfacets are of the (100) type~\cite{FEG07}, in agreement with 
our result. Threefold hexagonal islands of Au on O/Ru(111) have also been
observed upon annealing, evolving from dentritic shapes~\cite{Sch92}.

In homoepitaxial islands, the tensile stress present in unreconstructed
hexagonal surfaces~\cite{Iba94} should result in a hardly significant 
predominance of (100) steps according to our model. 
However, in these conditions, some more subtle effects such as those 
arising from the degree of $d$~band filling~\cite{Pap96} can be significant.
Nearly regular-hexagonal islands have been observed for Cu(111) and 
Ag(111)~\cite{Gie01B}. However, in the reconstructed Au(111) surface 
or in Pt(111), reconstructed at the temperatures and fluxes necessary 
to achieve growth of equilibrium shapes~\cite{Bott93}, the stress 
should be compressive and the predicted threefold hexagonal islands with
preferred (111)-faceted steps have indeed been observed~\cite{Mich93}.

Considering more complex systems beyond the applicability of our simple model,
triangular equilibrium shapes have been also observed for graphene islands
on Ni(111)~\cite{Li13} and WS$_2$ on Au(111)~\cite{Fuc13}.
Epitaxial graphene domains evolve towards triangular equilibrium shapes with only
one of the two possible zig-zag edge orientations, while for WS$_2$, triangles 
exposing either the so-called S- and W-edges have been observed. Epitaxial strain 
in both systems is very small. The reason for an energetic preference of one step 
type is not evident and interdiffusion of Ni and S adsorption on the edges
of WS$_2$ have been proposed, respectively. 
An overview of the experimental results discussed above that can be reasonably 
compared with the results of our simulations is given in Table~\ref{tbshapes}.
Although quantitative comparison is difficult in several cases, results 
are in reasonable agreement with the predictions of our model.

In summary, we have found that the step energies of strained hexagonal
monolayer islands depend on the amount of strain induced by the lattice
misfit. Edge energies decrease with increasing absolute value of the misfit due to 
strain relaxation. The effect is much more pronounced in the case of positive
misfit owing to the stronger repulsive interatomic forces. The energy of
(111)-faceted steps is lower at positive misfits (compressed islands) that of 
(100)-faceted steps. The ratio is reversed at negative misfits (tensile islands).
The equilibrium shape transforms from regular hexagons at zero misfit to islands
with threefold symmetric hexagonal shape with increasing misfit.
The results can be understood in terms of the different bonding geometries at the 
two different types of steps and have been found to compare reasonably with available
experiments.

This work was supported by projects FIS2008-01431 and FIS2011-23230 of
the Spanish MiCInn.

\end{document}